\pdfoutput=1
\documentclass[12pt,a4paper]{article}

\usepackage{ifthen}
\newboolean{pdflatex}
\setboolean{pdflatex}{true}

\newboolean{articletitles}
\setboolean{articletitles}{true}

\newboolean{uprightparticles}
\setboolean{uprightparticles}{false}

\newboolean{inbibliography}
\setboolean{inbibliography}{false}

\usepackage{rotating}

\usepackage{microtype}
\usepackage{lineno}
\usepackage{xspace}

\usepackage{multirow}
\usepackage{caption}

\usepackage{graphicx}
\usepackage{color}
\usepackage{colortbl}
\graphicspath{{./figs/}}

\usepackage{amsmath}
\usepackage{amssymb}
\usepackage{amsfonts}
\usepackage{upgreek}

\newcommand*\patchAmsMathEnvironmentForLineno[1]{
\expandafter\let\csname old#1\expandafter\endcsname\csname #1\endcsname
\expandafter\let\csname oldend#1\expandafter\endcsname\csname
end#1\endcsname
 \renewenvironment{#1}
   {\linenomath\csname old#1\endcsname}
   {\csname oldend#1\endcsname\endlinenomath}
}
\newcommand*\patchBothAmsMathEnvironmentsForLineno[1]{
  \patchAmsMathEnvironmentForLineno{#1}
  \patchAmsMathEnvironmentForLineno{#1*}
}
\AtBeginDocument{
\patchBothAmsMathEnvironmentsForLineno{equation}
\patchBothAmsMathEnvironmentsForLineno{align}
\patchBothAmsMathEnvironmentsForLineno{flalign}
\patchBothAmsMathEnvironmentsForLineno{alignat}
\patchBothAmsMathEnvironmentsForLineno{gather}
\patchBothAmsMathEnvironmentsForLineno{multline}
\patchBothAmsMathEnvironmentsForLineno{eqnarray}
}

\usepackage{hyperref}
\usepackage[all]{hypcap}

\def\lhcb {\mbox{LHCb}\xspace}

\ifthenelse{\boolean{uprightparticles}}
{

 \def\Ppi         {\ensuremath{\uppi}\xspace}

 \def\Ppsi        {\ensuremath{\uppsi}\xspace}

 \def\PDelta      {\ensuremath{\Delta}\xspace}
 \def\PXi      {\ensuremath{\Xi}\xspace}
 \def\PLambda      {\ensuremath{\Lambda}\xspace}
 \def\PSigma      {\ensuremath{\Sigma}\xspace}
 \def\POmega      {\ensuremath{\Omega}\xspace}
 \def\PUpsilon      {\ensuremath{\Upsilon}\xspace}

 \def\PB      {\ensuremath{\mathrm{B}}\xspace}
 
 \def\PD      {\ensuremath{\mathrm{D}}\xspace}

 \def\PJ      {\ensuremath{\mathrm{J}}\xspace}
 \def\PK      {\ensuremath{\mathrm{K}}\xspace}

 \def\Pb      {\ensuremath{\mathrm{b}}\xspace}
 \def\Pc      {\ensuremath{\mathrm{c}}\xspace}

 \def\Pi      {\ensuremath{\mathrm{i}}\xspace}

 \def\Pp      {\ensuremath{\mathrm{p}}\xspace}

 \def\Ps      {\ensuremath{\mathrm{s}}\xspace}

}
{

 \def\Ppi         {\ensuremath{\pi}\xspace}

 \def\Ppsi        {\ensuremath{\psi}\xspace}
 
 \mathchardef\PDelta="7101
 \mathchardef\PXi="7104
 \mathchardef\PLambda="7103
 \mathchardef\PSigma="7106
 \mathchardef\POmega="710A
 \mathchardef\PUpsilon="7107
 
 \def\PB      {\ensuremath{B}\xspace}
 
 \def\PD      {\ensuremath{D}\xspace}

 \def\PJ      {\ensuremath{J}\xspace}
 \def\PK      {\ensuremath{K}\xspace}

 \def\Pb      {\ensuremath{b}\xspace}
 \def\Pc      {\ensuremath{c}\xspace}

 \def\Pi      {\ensuremath{i}\xspace}

 \def\Pp      {\ensuremath{p}\xspace}

 \def\Ps      {\ensuremath{s}\xspace}

}

\makeatletter
\ifcase \@ptsize \relax
  \newcommand{\miniscule}{\@setfontsize\miniscule{4}{5}}
\or
  \newcommand{\miniscule}{\@setfontsize\miniscule{5}{6}}
\or
  \newcommand{\miniscule}{\@setfontsize\miniscule{5}{6}}
\fi
\makeatother

\DeclareRobustCommand{\optbar}[1]{\shortstack{{\miniscule (\rule[.5ex]{1.25em}{.18mm})}
  \\ [-.7ex] $#1$}}

\def\squark    {{\ensuremath{\Ps}}\xspace}

\def\cquark    {{\ensuremath{\Pc}}\xspace}

\def\bquark    {{\ensuremath{\Pb}}\xspace}

\def\pion   {{\ensuremath{\Ppi}}\xspace}
\def\piz    {{\ensuremath{\pion^0}}\xspace}

\def\pip    {{\ensuremath{\pion^+}}\xspace}
\def\pim    {{\ensuremath{\pion^-}}\xspace}

\def\kaon    {{\ensuremath{\PK}}\xspace}
  \def\Kbar    {{\kern 0.2em\overline{\kern -0.2em \PK}{}}\xspace}

\def\KorKbar    {\kern 0.18em\optbar{\kern -0.18em K}{}\xspace}

\def\Kp      {{\ensuremath{\kaon^+}}\xspace}
\def\Km      {{\ensuremath{\kaon^-}}\xspace}

\def\Kstarb  {{\ensuremath{\Kbar{}^*}}\xspace}

  \def\Dbar    {{\kern 0.2em\overline{\kern -0.2em \PD}{}}\xspace}
\def\D       {{\ensuremath{\PD}}\xspace}

\def\DorDbar    {\kern 0.18em\optbar{\kern -0.18em D}{}\xspace}
\def\Dz      {{\ensuremath{\D^0}}\xspace}
\def\Dzb     {{\ensuremath{\Dbar{}^0}}\xspace}

\def\Dstarzb {{\ensuremath{\Dbar{}^{*0}}}\xspace}

\def\Dstarpm {{\ensuremath{\D^{*\pm}}}\xspace}

\def\Dsm     {{\ensuremath{\D^-_\squark}}\xspace}

\def\B       {{\ensuremath{\PB}}\xspace}
\def\Bbar    {{\ensuremath{\kern 0.18em\overline{\kern -0.18em \PB}{}}}\xspace}

\def\BorBbar    {\kern 0.18em\optbar{\kern -0.18em B}{}\xspace}

\def\Bu      {{\ensuremath{\B^+}}\xspace}

\def\Bp      {{\ensuremath{\Bu}}\xspace}

\def\Bd      {{\ensuremath{\B^0}}\xspace}
\def\Bs      {{\ensuremath{\B^0_\squark}}\xspace}

\def\jpsi     {{\ensuremath{{\PJ\mskip -3mu/\mskip -2mu\Ppsi\mskip 2mu}}}\xspace}

  \def\Y#1S{\ensuremath{\PUpsilon{(#1S)}}\xspace}

\def\proton      {{\ensuremath{\Pp}}\xspace}
\def\antiproton  {{\ensuremath{\overline \proton}}\xspace}

\def\Lbar        {{\ensuremath{\kern 0.1em\overline{\kern -0.1em\PLambda}}}\xspace}
\def\LorLbar    {\kern 0.18em\optbar{\kern -0.18em \PLambda}{}\xspace}

\def\Lbbar   {{\ensuremath{\Lbar{}^0_\bquark}}\xspace}

\def\to                 {\ensuremath{\rightarrow}\xspace}

\def\AT#1     {\ensuremath{A_{\mathrm{T}}^{#1}}\xspace}

\def\C#1      {\ensuremath{\mathcal{C}_{#1}}\xspace}
\def\Cp#1     {\ensuremath{\mathcal{C}_{#1}^{'}}\xspace}
\def\Ceff#1   {\ensuremath{\mathcal{C}_{#1}^{\mathrm{(eff)}}}\xspace}
\def\Cpeff#1  {\ensuremath{\mathcal{C}_{#1}^{'\mathrm{(eff)}}}\xspace}
\def\Ope#1    {\ensuremath{\mathcal{O}_{#1}}\xspace}
\def\Opep#1   {\ensuremath{\mathcal{O}_{#1}^{'}}\xspace}

\newcommand{\tev}{\ifthenelse{\boolean{inbibliography}}{\ensuremath{~T\kern -0.05em eV}\xspace}{\ensuremath{\mathrm{\,Te\kern -0.1em V}}}\xspace}
\newcommand{\gev}{\ensuremath{\mathrm{\,Ge\kern -0.1em V}}\xspace}
\newcommand{\mev}{\ensuremath{\mathrm{\,Me\kern -0.1em V}}\xspace}
\newcommand{\kev}{\ensuremath{\mathrm{\,ke\kern -0.1em V}}\xspace}
\newcommand{\ev}{\ensuremath{\mathrm{\,e\kern -0.1em V}}\xspace}
\newcommand{\gevc}{\ensuremath{{\mathrm{\,Ge\kern -0.1em V\!/}c}}\xspace}
\newcommand{\mevc}{\ensuremath{{\mathrm{\,Me\kern -0.1em V\!/}c}}\xspace}
\newcommand{\gevcc}{\ensuremath{{\mathrm{\,Ge\kern -0.1em V\!/}c^2}}\xspace}
\newcommand{\gevgevcccc}{\ensuremath{{\mathrm{\,Ge\kern -0.1em V^2\!/}c^4}}\xspace}
\newcommand{\mevcc}{\ensuremath{{\mathrm{\,Me\kern -0.1em V\!/}c^2}}\xspace}

\def\mm   {\ensuremath{\rm \,mm}\xspace}

\newcommand{\chisq}{\ensuremath{\chi^2}\xspace}

\def\gsim{{~\raise.15em\hbox{$>$}\kern-.85em
          \lower.35em\hbox{$\sim$}~}\xspace}
\def\lsim{{~\raise.15em\hbox{$<$}\kern-.85em
          \lower.35em\hbox{$\sim$}~}\xspace}

\def\sPlot{\mbox{\em sPlot}}

\def\pt         {\mbox{$p_{\rm T}$}\xspace}

\def\tell1  {TELL1\xspace}
\def\ukl1   {UKL1\xspace}

\newcommand{\ie}{\mbox{\itshape i.e.}\xspace}

\newcommand{\vs}{\mbox{\itshape vs.}\xspace}

\def\Kstarbsubz  {{\ensuremath{\Kbar{}^*_0}}\xspace}

\def\Kstarbsubt  {{\ensuremath{\Kbar{}^*_2}}\xspace}

\def\DorDstarzb {{\ensuremath{\Dbar{}^{(*)0}}}\xspace}

\usepackage{cite}
\usepackage{mciteplus}

\begin{document}

\renewcommand{\thefootnote}{\fnsymbol{footnote}}
\setcounter{footnote}{1}

\begin{titlepage}
\pagenumbering{roman}

\vspace*{-1.5cm}
\centerline{\large EUROPEAN ORGANIZATION FOR NUCLEAR RESEARCH (CERN)}
\vspace*{1.5cm}
\hspace*{-0.5cm}
\begin{tabular*}{\linewidth}{lc@{\extracolsep{\fill}}r}
\ifthenelse{\boolean{pdflatex}}
{\vspace*{-2.7cm}\mbox{\!\!\!\includegraphics[width=.14\textwidth]{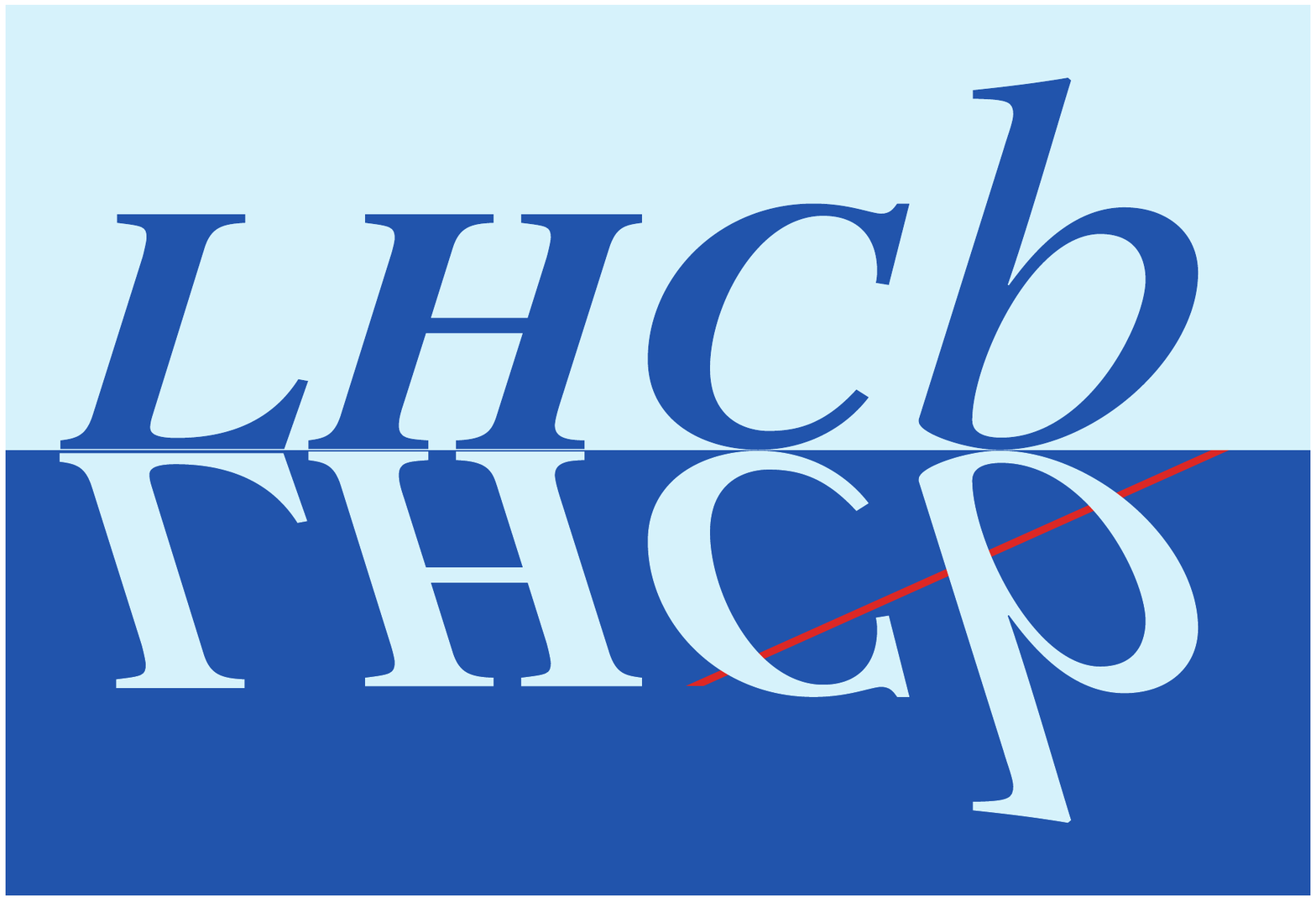}} & &}
{\vspace*{-1.2cm}\mbox{\!\!\!\includegraphics[width=.12\textwidth]{lhcb-logo.eps}} & &}
\\
 & & CERN-PH-EP-2014-183 \\
 & & LHCb-PAPER-2014-035 \\
 & & 20 October 2014 \\
 & & \\
\end{tabular*}

\vspace*{2.0cm}

{\bf\boldmath\huge
\begin{center}
  Observation of overlapping spin-1 and spin-3 $\Dzb\Km$ resonances at mass $2.86 \gevcc$
\end{center}
}

\vspace*{1.0cm}

\begin{center}
  The LHCb collaboration\footnote{Authors are listed at the end of this Letter.}
\end{center}

\vspace{\fill}

\begin{abstract}
  \noindent
  The resonant substructure of $\Bs \to \Dzb\Km\pip$ decays is studied using a data sample corresponding to an integrated luminosity of $3.0  \,{\rm fb}^{-1}$ of $pp$ collision data recorded by the LHCb detector.
  An excess at $m(\Dzb\Km) \approx 2.86 \gevcc$ is found to be an admixture of spin-1 and spin-3 resonances.
  Therefore the $D^{*}_{sJ}(2860)^{-}$ state previously observed in inclusive $e^+e^- \to \Dzb\Km X$ and $pp \to \Dzb\Km X$ processes consists of at least two particles.
  This is the first observation of a heavy flavoured spin-3 resonance, and the first time that any spin-3 particle has been seen to be produced in \B decays.
  The masses and widths of the new states and of the $D^{*}_{s2}(2573)^{-}$ meson are measured, giving the most precise determinations to date.
\end{abstract}

\vspace*{2.0cm}

\begin{center}
  Published in Phys.~Rev.~Lett.
\end{center}

\vspace{\fill}

{\footnotesize
\centerline{\copyright~CERN on behalf of the \lhcb collaboration, license \href{http://creativecommons.org/licenses/by/4.0/}{CC-BY-4.0}.}}
\vspace*{2mm}

\end{titlepage}

\newpage
\setcounter{page}{2}
\mbox{~}

\cleardoublepage

\renewcommand{\thefootnote}{\arabic{footnote}}
\setcounter{footnote}{0}

\pagestyle{plain}
\setcounter{page}{1}
\pagenumbering{arabic}

Studies of heavy meson spectroscopy provide an important probe of quantum chromodynamics.
The observations of the $D_{s0}^*(2317)^-$~\cite{Aubert:2003fg} and
$D_{s1}(2460)^-$~\cite{Besson:2003cp} mesons led to renewed activity in the
field, as their masses were found to be below the $DK$ and $D^*K$ thresholds,
respectively, in contrast to prior predictions.
These states are usually interpreted as being two of the orbitally excited
(1P) charm-strange states, the other two being the $D_{s1}(2536)^-$ and
$D_{s2}^*(2573)^-$ resonances.
Several other charm-strange states, the $D_{s1}^*(2700)^-$, $D_{sJ}^*(2860)^-$ and $D_{sJ}(3040)^-$ resonances, have been discovered~\cite{Aubert:2006mh,Brodzicka:2007aa,Aubert:2009ah,LHCb-PAPER-2012-016}.
However, their quantum numbers and spectroscopic assignments are not known, with the exception of the $D_{s1}^*(2700)^-$ meson, which has spin-parity $J^P = 1^-$ and is generally believed to be a radially excited (2S) state.
Reviews of the expectations in theoretical models can be found in Refs.~\cite{Swanson:2006st,Rosner:2006jz,Klempt:2007cp,Colangelo:2012xi}.

A state with $J^P = 3^-$ would be a clear candidate for a member of the 1D family, \ie\ a state with two units of orbital excitation.
Spin-3 states have been observed in the light unflavoured~\cite{Wagner:1974gw,Aston:1988rf} and strange~\cite{Brandenburg:1975ft,Baldi:1976ua} meson sectors, but not previously among heavy flavoured mesons.
Production of high-spin states is expected to be suppressed in \B meson decays, and has not previously been observed~\cite{blatt-weisskopf}.
However, high-spin resonances are expected to be relatively narrow, potentially enhancing their observability.

Analysis of the Dalitz plot~\cite{Dalitz:1953cp} that describes the phase-space of a three-body decay is a powerful tool for spectroscopic studies.
Compared to measurements based on inclusive production processes, the lower
background level allows broader states to be distinguished and the well-defined initial state allows the quantum numbers to be unambiguously determined.
Specifically, in $\Bs\to\Dzb\Km\pip$ decays, $\Km\pip$ and $\Dzb\Km$ resonances appear as horizontal and vertical bands in the Dalitz plot formed from the invariant masses squared $m^2(\Km\pip)$ \vs\ $m^2(\Dzb\Km)$, and the spin of the resonance can be inferred from the distribution of decays along the band.
Measurement of the spin also determines the parity, since only natural spin-parity resonances can decay strongly to two pseudoscalars.

In this Letter, results of the first Dalitz plot analysis of the $\Bs \to \Dzb \Km \pip$ decay are summarised.
The inclusion of charge conjugated processes is implied throughout the paper.
The \Dzb meson is reconstructed through the $\Kp\pim$ decay mode, which is
treated as flavour-specific, \ie\ the heavily suppressed $\Bs \to \Dz \Km \pip, \Dz \to \Kp\pim$ contribution is neglected.
The amplitude analysis technique is used to separate contributions from excited charm-strange mesons and from excited kaon states.
A detailed description of the analysis can be found in Ref.~\cite{LHCb-PAPER-2014-036}.

The analysis is based on a data sample corresponding to $3.0  \,{\rm fb}^{-1}$ of integrated luminosity, approximately one third (two thirds) of which was collected by the LHCb detector from $pp$ collisions at a centre-of-mass energy of $7 \ (8) \ \tev$ during 2011 (2012).
The \lhcb detector is a single-arm forward spectrometer covering the \mbox{pseudorapidity} range $2<\eta <5$, designed for the study of particles containing \bquark or \cquark quarks, and is described in detail in Ref.~\cite{Alves:2008zz}.
Signal candidates are accepted offline if one of the final state particles deposited sufficient energy transverse to the beamline in the hadronic calorimeter to fire the hardware trigger.
Events that are triggered at the hardware level by another particle in the event are also retained.
The software trigger~\cite{LHCb-DP-2012-004} requires a two-, three- or four-track secondary vertex with a large sum of the transverse momentum, \pt, of the tracks and a significant displacement from all primary $pp$ interaction vertices~(PVs).

The offline selection requirements are similar to those used in Refs.~\cite{LHCb-PAPER-2012-056,LHCb-PAPER-2013-022} and are optimised using the $\Bd\to\Dzb\pip\pim$ decay as a control channel.
Discrimination between signal and background categories is achieved primarily with a neural network~\cite{Feindt:2006pm} trained on $\Bd\to\Dzb\pip\pim$ data, where signal and background are statistically separated with the \sPlot\ technique~\cite{Pivk:2004ty} using the \B candidate mass as discriminating variable.
A total of 16 variables are used in the network.
They include the output of a ``\Dz boosted decision tree''~\cite{LHCb-PAPER-2012-025,LHCb-PAPER-2012-050} that identifies \Dzb mesons produced in \bquark hadron decays, together with other variables that characterise the topology and the kinematic distributions of the \B decay.
A requirement on the network output is imposed that reduces the combinatorial background remaining after the initial selection by a factor of five while retaining more than $90\,\%$ of the signal.
The four final state tracks also have to satisfy pion and kaon identification requirements.

To improve the mass resolution, track momenta are scaled~\cite{LHCb-PAPER-2012-048,LHCB-PAPER-2013-011} with calibration parameters determined by
matching the observed position of the dimuon mass peak to the known $\jpsi$ mass~\cite{PDG2012}.
In addition, the momenta of the tracks from the $\Dzb$ candidate are adjusted~\cite{Hulsbergen:2005pu} so that their combined invariant mass matches the known $\Dzb$ mass~\cite{PDG2012}.
An additional $\Bs$ mass constraint is applied in the calculation of the Dalitz plot variables.

Invariant-mass vetoes are applied to remove backgrounds containing $\Dstarpm$ mesons, and from the $\Bs\to\Dsm\pip$ and $\Bs \to \Dz\Dzb$ decays.
Decays of \Bs mesons to the same final state but without an intermediate charm
meson are suppressed by the $\Dz$ boosted decision tree criteria and an additional requirement that the $\Dzb$ candidate vertex is displaced by at least $1\mm$ from the $\Bs$ decay vertex.

The signal and background yields are obtained from an extended unbinned maximum
likelihood fit to the three-body invariant mass distribution of $\Bs \to \Dzb \Km\pip$ candidates in the range $5200$--$5900 \mevcc$.
In addition to signal decays and combinatorial background, the fit model includes components to describe: partially reconstructed $\Bs \to \Dstarzb \Km\pip$ decays, with $\Dstarzb \to \Dzb\piz$ or $\Dzb\gamma$ and the $\piz$ or $\gamma$ not included in the reconstruction; $\Bd \to \Dzb \Km\pip$ decays; and $\Bd \to \DorDstarzb \pip\pim$ and $\Lbbar \to \DorDstarzb \antiproton \pip$~\cite{LHCb-PAPER-2013-056} decays with misidentification of a final state particle.
Contributions from other $\Bs$ and $\Bd$ decays are negligible.

The signal and $\Bd \to \Dzb \Km\pip$ shapes are each modelled with the sum of two Crystal Ball~\cite{Skwarnicki:1986xj} functions which share a common mean and have tails on opposite sides.
The combinatorial background is modelled using a linear function.
Smoothed histograms are used to describe the shapes of $\Bs \to \Dstarzb
\Km\pip$, $\Bd \to \DorDstarzb \pip\pim$ and $\Lbbar \to \DorDstarzb
\antiproton \pip$ decays.
These shapes are determined from simulated events reweighted to account for the known Dalitz plot distributions of the background decays~\cite{LHCb-PAPER-2013-022,LHCb-PAPER-2013-056} and particle identification and misidentification probabilities.

\begin{figure}[!tb]
  \centering
  \includegraphics[scale=0.50]{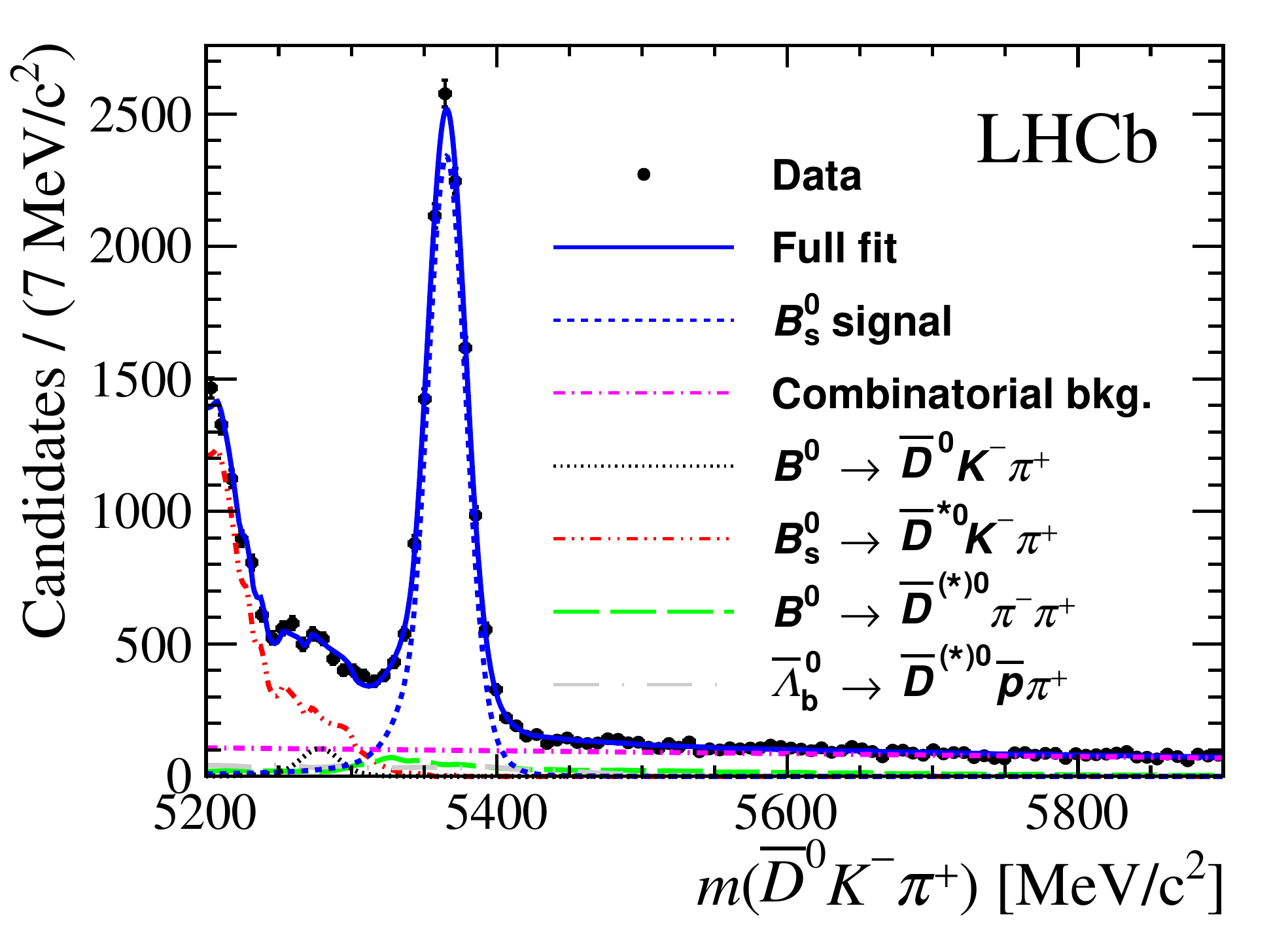}
  \caption{\small
    Fit to the $\Bs\to \Dzb\Km\pip$ candidate invariant mass distribution.
    Data points are shown in black, the result of the fit as a solid blue line and the components as detailed in the legend.
  }
  \label{fig:fits}
\end{figure}

The results of the fit are shown in Fig.~\ref{fig:fits}.
Within a signal region of $\mu_\Bs\pm2.5\sigma_1$, where the peak position $\mu_\Bs$ and core width $\sigma_{1} = 12.7 \pm 0.2 \mevcc$ are taken from the results of the fit, there are $12\,954$ candidates.
Of these, $11\,300\pm160$ are signal decays, while $950\pm60$ are combinatorial background, $360\pm130$ are $\Bd\to\DorDstarzb \pip\pim$ decays and $300\pm80$ are $\Lbbar \to \DorDstarzb \antiproton\pip$ decays.

\begin{figure}[!tb]
  \centering
  \includegraphics[scale=0.50]{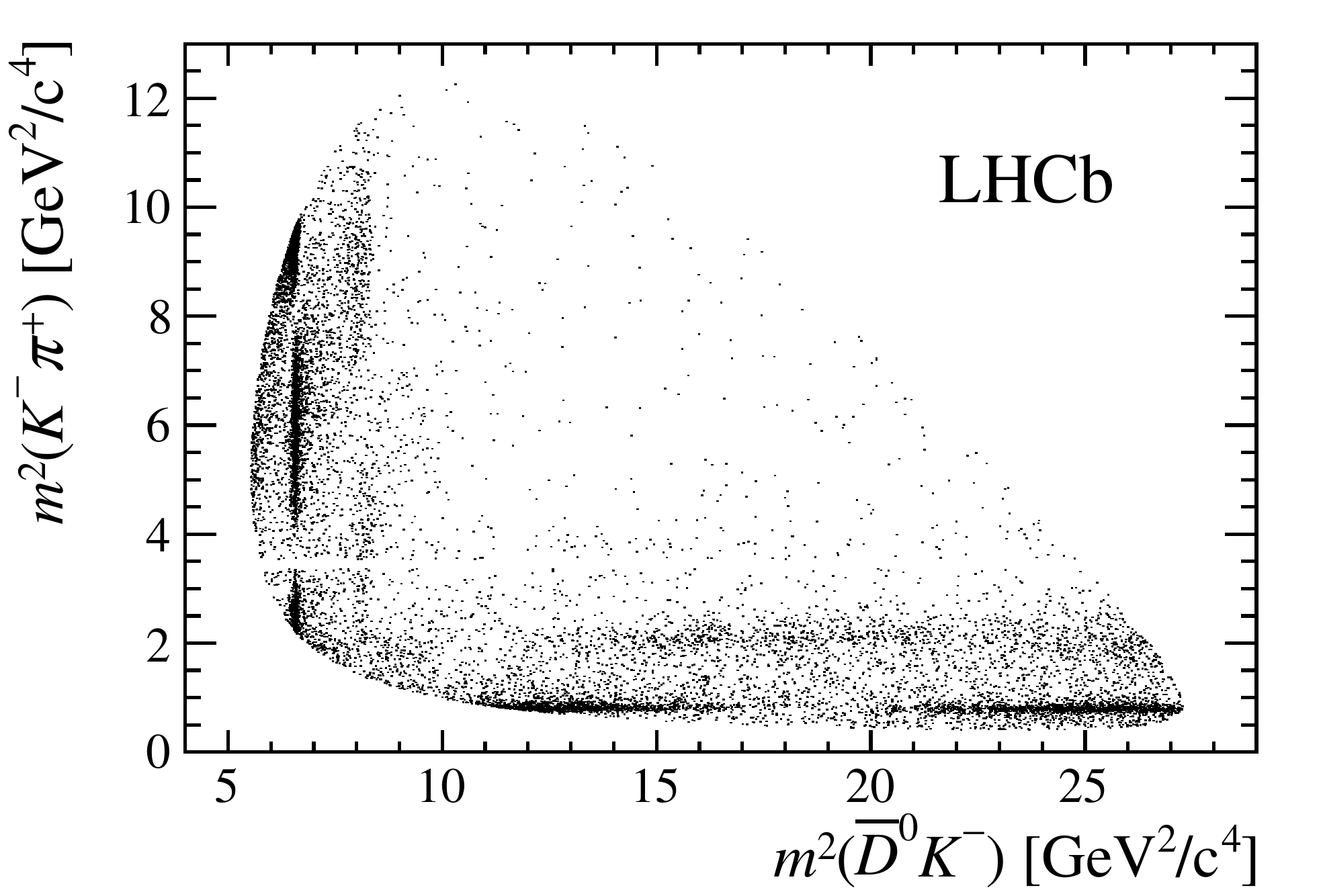}
  \caption{\small
    Dalitz plot distribution of $\Bs\to \Dzb\Km\pip$ candidates in the signal region.
    The effect of the \Dz veto can be seen as an unpopulated horizontal band.
  }
  \label{fig:signalevents}
\end{figure}

The Dalitz plot distribution of the candidates in the signal region, shown in Fig.~\ref{fig:signalevents}, is fitted with a model that includes both signal and background components.
The Dalitz plot distribution of combinatorial background is obtained from a sideband region above the signal peak in the \Bs candidate mass, while those for $\Bd\to\DorDstarzb \pip\pim$ and $\Lbbar \to \DorDstarzb \antiproton\pip$ backgrounds are obtained from simulation reweighted in the same way as their \Bs candidate mass shapes.

The signal model is defined by considering many possible contributions and removing those that do not significantly affect the fit.
It contains 15 resonant or nonresonant amplitudes added coherently in the isobar~\cite{Fleming:1964zz,Morgan:1968zza,Herndon:1973yn} model formalism.
These include the $\Kstarb(892)^{0}$, $\Kstarb(1410)^{0}$, $\Kstarbsubt(1430)^{0}$ and $\Kstarb(1680)^{0}$ resonances.
The $\Km\pip$ S-wave is modelled using the LASS shape~\cite{lass}, which combines the $\Kstarbsubz(1430)^{0}$ resonance with a slowly varying (nonresonant) component, in addition to the $\Kstarbsubz(1950)^{0}$ state.
The $D^{*}_{s2}(2573)^{-}$ and $D^{*}_{s1}(2700)^{-}$ states are included, in addition to both spin-1 and spin-3 resonances near $m(\Dzb\Km) \approx 2860 \mevcc$ labelled $D^{*}_{s1}(2860)^{-}$ and $D^{*}_{s3}(2860)^{-}$, respectively.
A nonresonant S-wave $\Dzb\Km$ component is included, modelled with an exponential form factor, as are additional amplitudes mediated by ``virtual'' resonances (\ie\ with peak position outside the phase space of the Dalitz plot so that only the tail of the lineshape contributes):  $D^{*-}_{s\,v}$ and $D^{*}_{s0\,v}(2317)^{-}$ in $m(\Dzb\Km)$, and $B^{*+}_{v}$ in $m(\Dzb\pip)$.
All components, except those explicitly noted above, are modelled with relativistic Breit--Wigner functions.
The parameters of the lineshapes are fixed to their known values~\cite{PDG2012}, except for the masses and widths of the $D^{*}_{s2}(2573)^{-}$, $D^{*}_{s1}(2860)^{-}$ and $D^{*}_{s3}(2860)^{-}$ resonances, the parameters describing the LASS function and the exponential form factor of the nonresonant model, which are free to vary in the fit.
The angular distributions are given in the Zemach tensor formalism~\cite{Zemach:1963bc,Zemach:1968zz} and each amplitude includes Blatt--Weisskopf barrier form factors~\cite{blatt-weisskopf}.

The signal model is multiplied by an efficiency function and normalised to unity when integrated across the Dalitz plot.
The efficiency is determined as a function of Dalitz plot position from samples of simulated events with corrections applied for known discrepancies between data and simulation in the efficiencies of the trigger, track reconstruction and particle identification.
The trigger efficiency correction is applied separately for candidates in events triggered at hardware level by the signal decay products and for those triggered independently.
The largest source of efficiency variation across the Dalitz plot arises due to a rapid decrease of the probability to reconstruct low momentum particles.
The particle identification requirements lead to a maximum efficiency variation of about $\pm 20\,\%$, while other effects are smaller.

\begin{figure}[!tb]
\centering
 \includegraphics[scale=0.39]{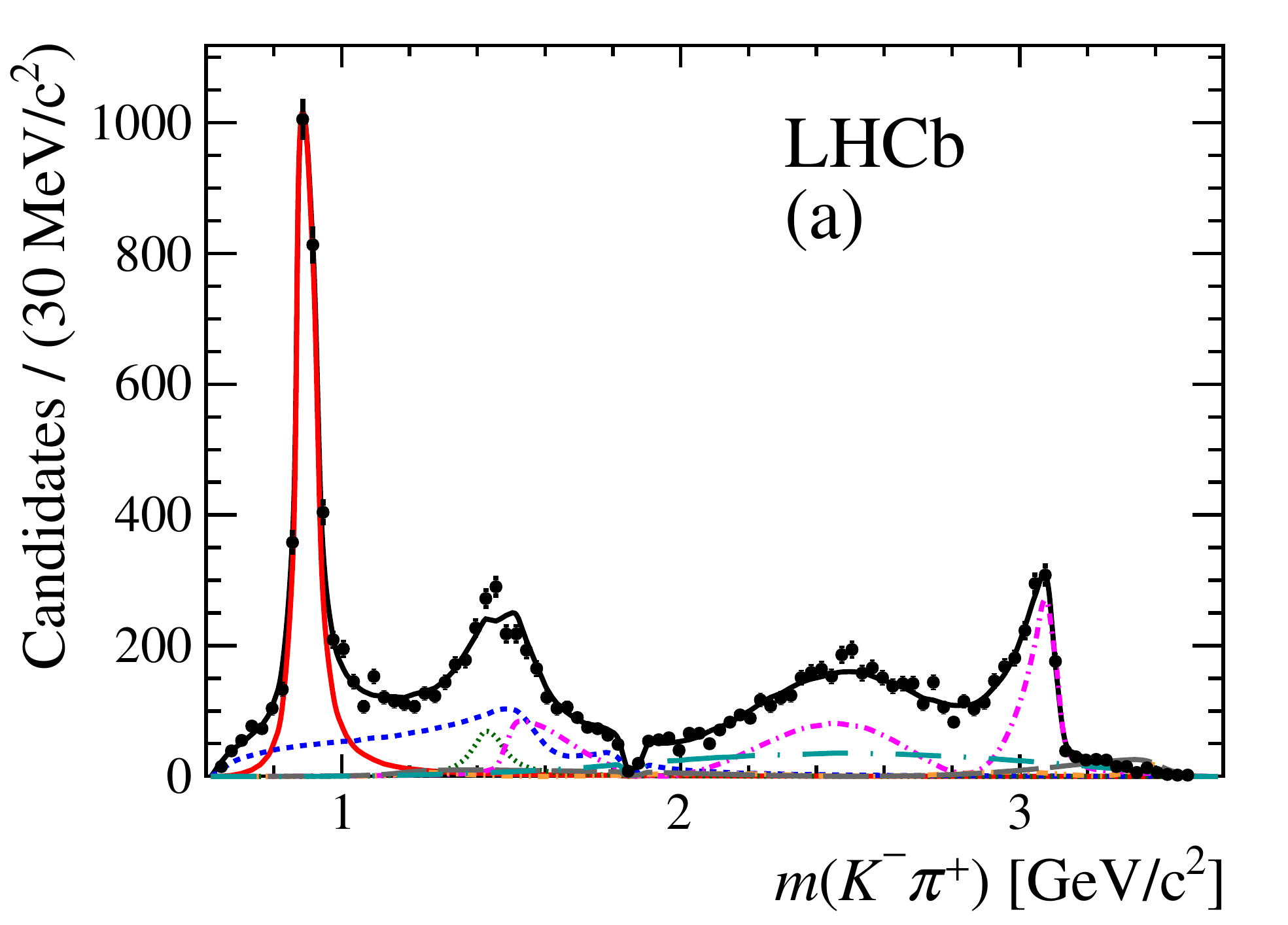}
 \includegraphics[scale=0.39]{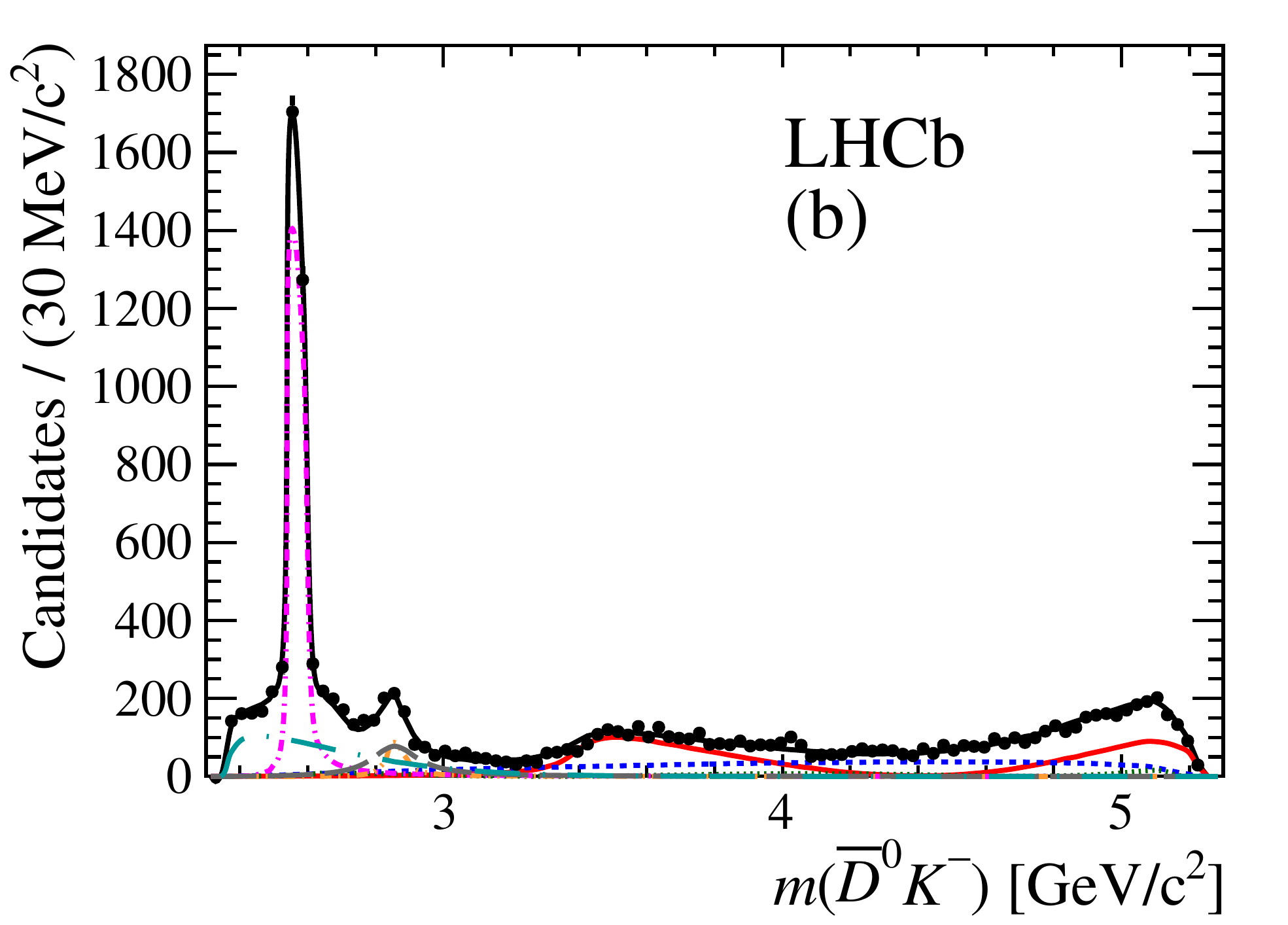}
 \includegraphics[scale=0.39]{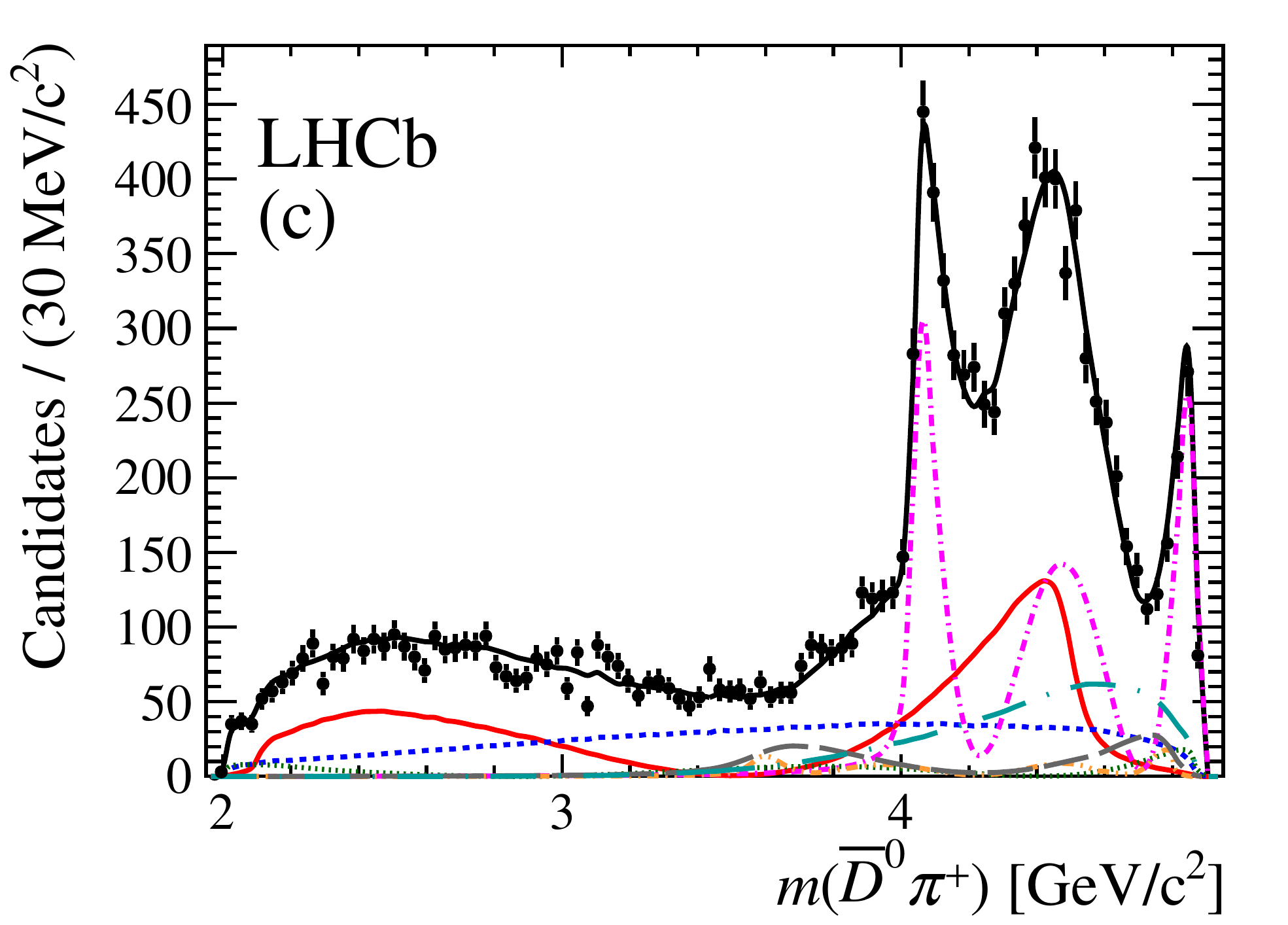}
 \includegraphics[scale=0.39]{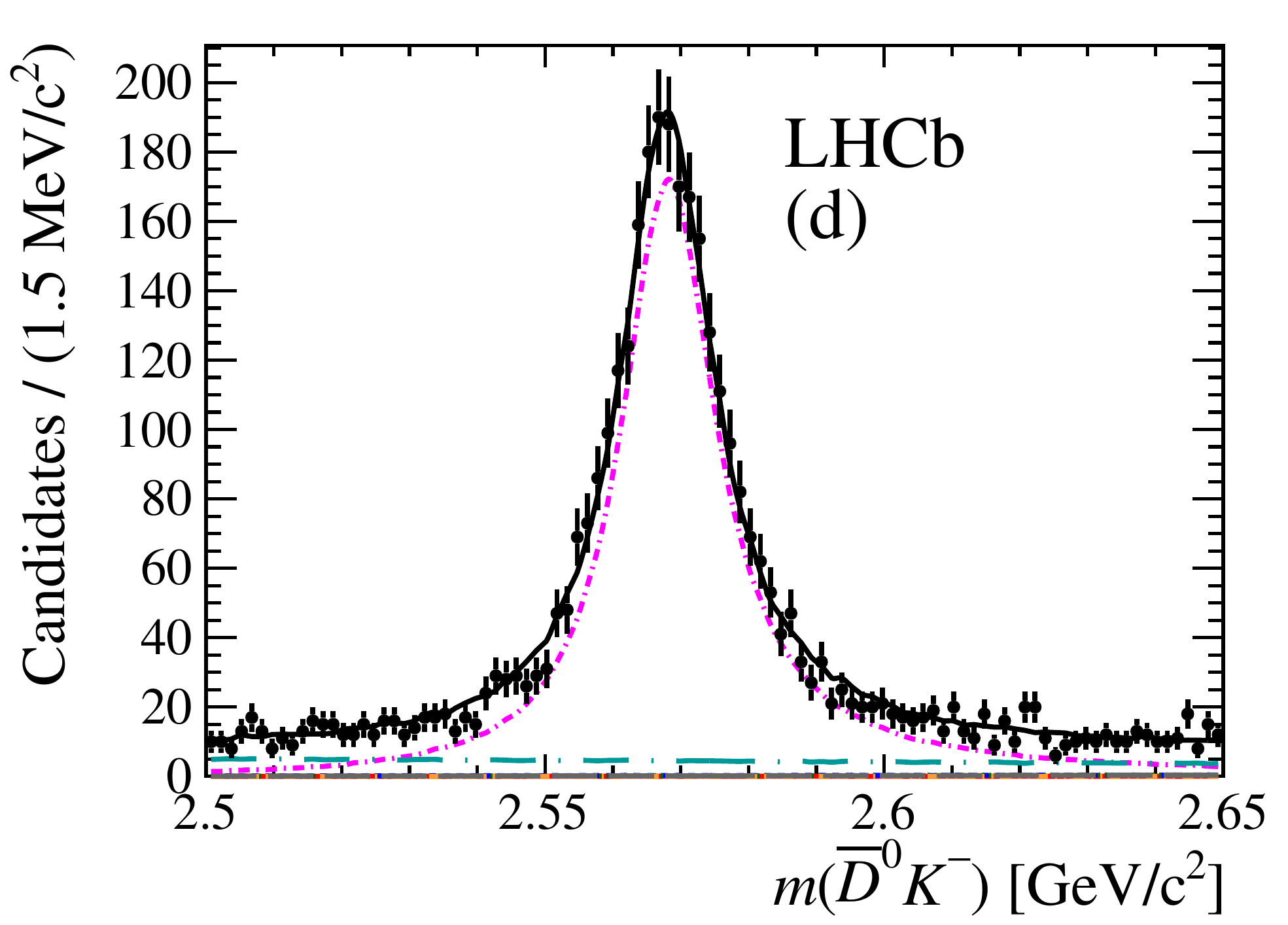}
 \includegraphics[scale=0.39]{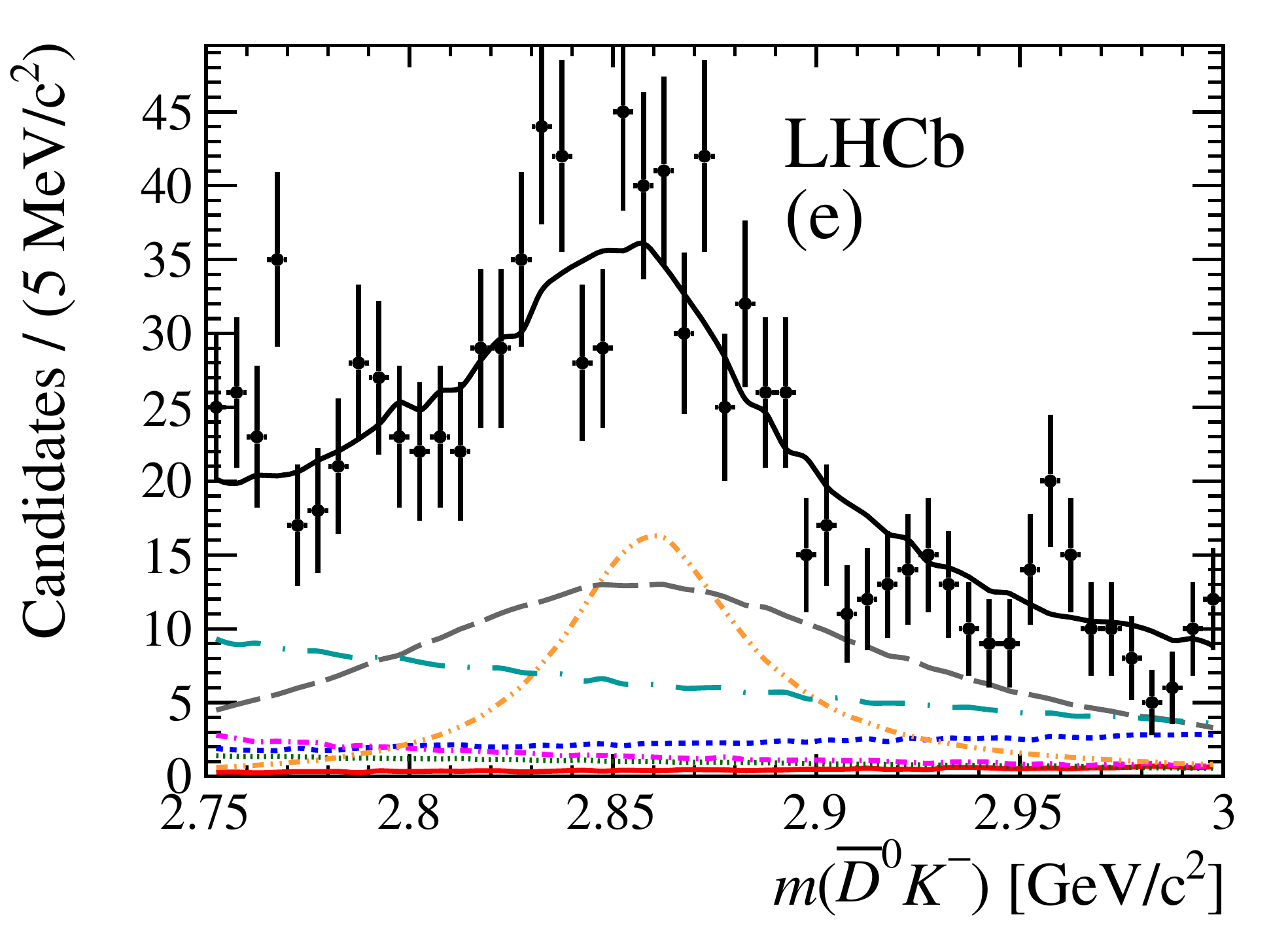}
 \includegraphics[scale=0.39]{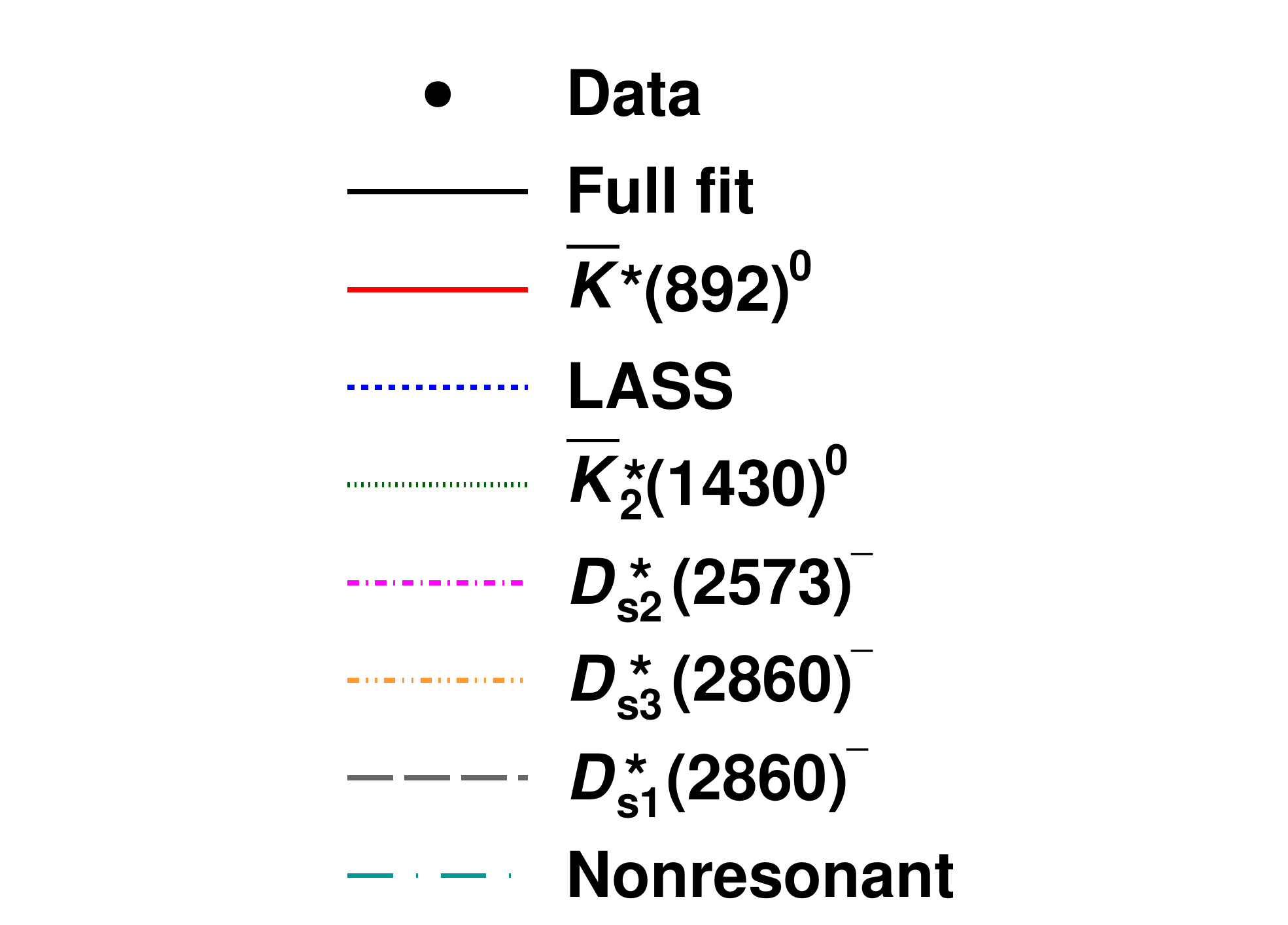}
\caption{\small
  Projections of the data and the Dalitz plot fit result onto
  (a) $m(\Km\pip)$, (b) $m(\Dzb\Km)$ and (c) $m(\Dzb\pip)$, with zooms into
  $m(\Dzb\Km)$ around (d) the $D^{*}_{s2}(2573)^{-}$ resonance and (e) the $D^{*}_{sJ}(2860)^{-}$ region.
  The data are shown as black points, the total fit result as a solid black
  curve, and the contributions from different resonances as detailed in the legend (small components, including the background contributions, are not shown).
}
\label{fig:DPprojections}
\end{figure}

Projections of the data and the unbinned maximum likelihood fit result are shown in Fig.~\ref{fig:DPprojections}.
The largest components in terms of their fit fractions, defined as the ratio of the integrals over the Dalitz plot of a single decay amplitude squared and the total amplitude squared, are the $\Kstarb(892)^{0}$ ($28.6\,\%$), $D^{*}_{s2}(2573)^{-}$ ($25.7\,\%$), LASS ($21.4\,\%$) and $\Dzb\Km$ nonresonant ($12.4\,\%$) terms.
The fit fractions for the $D^{*}_{s1}(2860)^{-}$ and $D^{*}_{s3}(2860)^{-}$ components are $(5.0 \pm 1.2 \pm 0.7 \pm 3.3)\,\%$ and $(2.2 \pm 0.1 \pm 0.3 \pm 0.4)\,\%$, respectively, where the uncertainties are statistical, systematic and from Dalitz plot model variations, as described below.
The phase difference between the $D^{*}_{s1}(2860)^{-}$ and $D^{*}_{s3}(2860)^{-}$ amplitudes is consistent with $\pi$ within a large model uncertainty.

\begin{figure}[!tb]
\centering
\includegraphics[scale=0.50]{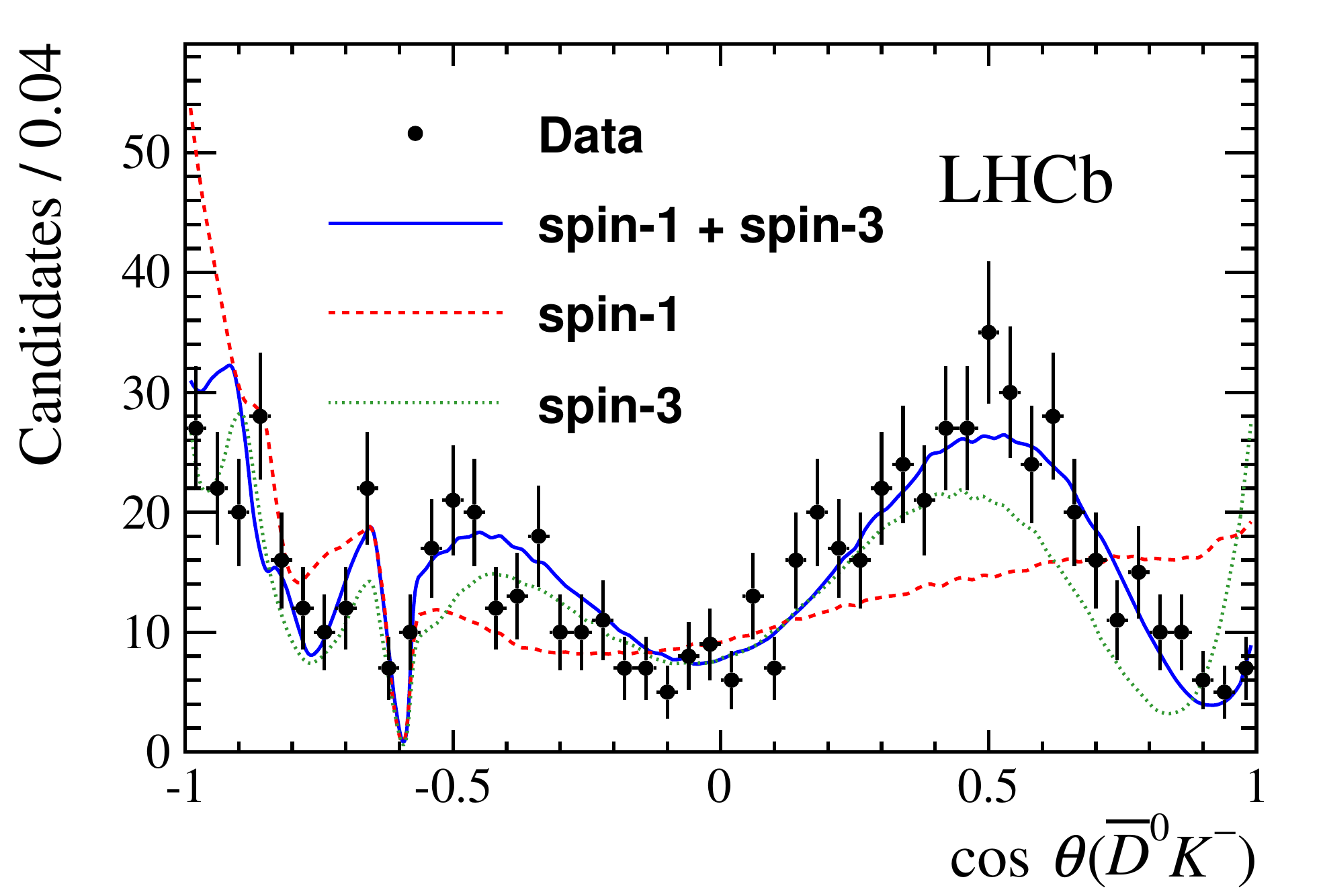}
\caption{\small
  Projections of the data and Dalitz plot fit results with alternative models
  onto the cosine of the helicity angle of the $\Dzb\Km$ system,
  $\cos\theta(\Dzb\Km)$, for $2.77 < m(\Dzb\Km) < 2.91\gevcc$,
  where $\theta(\Dzb\Km)$ is the angle between the $\pip$ and the $\Dzb$ meson momenta in the $\Dzb\Km$ rest frame.
    The data are shown as black points, with the fit results with different models as detailed in the legend.
  The dip at $\cos\theta(\Dzb\Km) \approx -0.6$ is due to the $\Dzb$ veto.
  Comparisons of the data and the different fit results in the 50 bins of this
  projection give $\chisq$ values of 47.3, 214.0 and 150.0 for the default,
  spin-1 only and spin-3 only models, respectively.
}
\label{fig:dsjsep}
\end{figure}

To assess the significance of the two states near $m(\Dzb\Km) \approx 2860 \mevcc$, the fit is repeated with all combinations of either one or two resonant amplitudes with different spins up to and including 3.
All other combinations give values of negative log-likelihood more than one hundred units larger than the default fit.
A comparison of the angular distributions in the region near $m(\Dzb\Km) \approx 2860 \mevcc$ of the data and the best fits with the spin-1 only, spin-3 only and both resonances is presented in Fig.~\ref{fig:dsjsep}.
Including both spin components visibly improves the fit.
Large samples of pseudoexperiments are generated with signal models corresponding to the best fits with the spin-1 or spin-3 amplitude removed, and each pseudoexperiment is fitted under both the one- and two-resonance hypotheses.
By extrapolating the tails of the distributions of the difference in negative log-likelihood values to the values observed in data, the statistical significances of the spin-3 and spin-1 components are found to be $16$ and $15$ standard deviations, respectively.
These significances remain in excess of 10 standard deviations in all alternative models considered below.

The considered sources of systematic uncertainty are divided into two main categories: experimental uncertainties and model uncertainties.
The experimental systematic uncertainties arise from imperfect knowledge of:
the relative amount of signal and background in the selected events;
the distributions of each of the backgrounds across the phase space;
the variation of the efficiency across the phase space;
the possible bias induced by the fit procedure;
the momentum calibration;
the fixed masses of the $\Bs$ and $\Dzb$ mesons used to define the boundaries of the Dalitz plot.
Model uncertainties occur due to:
fixed parameters in the Dalitz plot model;
the treatment of marginal components in the default fit model;
the choice of models for the $\Km\pip$ S-wave, the $\Dzb\Km$ S- and P-waves, and the lineshapes of the virtual resonances.
The systematic uncertainties from each source are combined in quadrature.

The masses and widths of the $D^{*}_{s2}(2573)^{-}$, $D^{*}_{s1}(2860)^{-}$ and $D^{*}_{s3}(2860)^{-}$ states are determined to be
\begin{eqnarray*}
  m(D^{*}_{s2}(2573)^-)      & = & 2568.39 \pm 0.29 \pm 0.19 \pm 0.18\mevcc \, ,\\
  \Gamma(D^{*}_{s2}(2573)^-) & = & 16.9    \pm 0.5  \pm 0.4  \pm 0.4 \mevcc \, ,\\
  m(D^{*}_{s1}(2860)^-)      & = & 2859    \pm 12   \pm 6    \pm 23  \mevcc \, ,\\
  \Gamma(D^{*}_{s1}(2860)^-) & = & 159      \pm 23   \pm 27   \pm 72  \mevcc \, ,\\
  m(D^{*}_{s3}(2860)^-)      & = & 2860.5   \pm 2.6  \pm 2.5  \pm 6.0 \mevcc \, ,\\
  \Gamma(D^{*}_{s3}(2860)^-) & = & 53       \pm 7    \pm 4    \pm 6   \mevcc \, ,
\end{eqnarray*}
where the first uncertainty is statistical, the second is due to experimental systematic effects and the third due to model variations.
The largest sources of uncertainty on the parameters of the $D^{*}_{s1}(2860)^-$ and $D^{*}_{s3}(2860)^-$ resonances arise from varying the $\Km\pip$ S-wave description and, for the $D^{*}_{s1}(2860)^{-}$ width, from removing the $\Kstarb(1680)^{0}$ and $B^{*+}_{v}$ components from the model.
The results for the $D^{*}_{s2}(2573)^-$ mass and width are determined with significantly better precision than previous measurements.
Those for the parameters of the $D^{*}_{s1}(2860)^-$ and $D^{*}_{s3}(2860)^-$ resonances must be considered first measurements, since previous measurements of the properties of the $D^{*}_{sJ}(2860)^-$ state~\cite{Aubert:2006mh,Aubert:2009ah,LHCb-PAPER-2012-016} involved an unknown admixture of at least these two particles.
The results for all the complex amplitudes determined by the Dalitz plot fit, as well as derived quantities such as branching fractions of the resonant contributions and detailed descriptions of the systematic uncertainties, are given in Ref.~\cite{LHCb-PAPER-2014-036}.

In summary, results of the first amplitude analysis of the $\Bs \to \Dzb \Km \pip$ decay show, with significance of more than 10 standard deviations, that a structure at $m(\Dzb\Km) \approx 2.86 \gevcc$ contains both spin-1 and spin-3 components.
The masses of the $D_{s1}^*(2860)^-$ and $D_{s3}^*(2860)^{-}$ states are found to be similar, while a larger width of the spin-1 state than that of the spin-3 state is preferred.
The results support an interpretation of these states being the $J^P = 1^-$ and $3^-$ members of the 1D family, though the $1^-$ state may be partially mixed with the vector member of the 2S family to give the physical $D_{s1}^*(2700)^-$ and $D_{s1}^*(2860)^-$ states.
The discovery of the $D_{s3}^*(2860)^-$ resonance represents the first observation of a heavy flavoured spin-3 particle, and the first time that a spin-3 state is seen to be produced in $B$ decays.
This demonstrates that the spectroscopy of the 1D families of heavy flavoured mesons can be studied experimentally.

\section*{Acknowledgements}

\noindent We express our gratitude to our colleagues in the CERN
accelerator departments for the excellent performance of the LHC. We
thank the technical and administrative staff at the LHCb
institutes. We acknowledge support from CERN and from the national
agencies: CAPES, CNPq, FAPERJ and FINEP (Brazil); NSFC (China);
CNRS/IN2P3 (France); BMBF, DFG, HGF and MPG (Germany); SFI (Ireland); INFN (Italy);
FOM and NWO (The Netherlands); MNiSW and NCN (Poland); MEN/IFA (Romania);
MinES and FANO (Russia); MinECo (Spain); SNSF and SER (Switzerland);
NASU (Ukraine); STFC (United Kingdom); NSF (USA).
The Tier1 computing centres are supported by IN2P3 (France), KIT and BMBF
(Germany), INFN (Italy), NWO and SURF (The Netherlands), PIC (Spain), GridPP
(United Kingdom).
We are indebted to the communities behind the multiple open
source software packages on which we depend. We are also thankful for the
computing resources and the access to software R\&D tools provided by Yandex LLC (Russia).
Individual groups or members have received support from
EPLANET, Marie Sk\l{}odowska-Curie Actions and ERC (European Union),
Conseil g\'{e}n\'{e}ral de Haute-Savoie, Labex ENIGMASS and OCEVU,
R\'{e}gion Auvergne (France), RFBR (Russia), XuntaGal and GENCAT (Spain), Royal Society and Royal
Commission for the Exhibition of 1851 (United Kingdom).

\ifx\mcitethebibliography\mciteundefinedmacro
\PackageError{LHCb.bst}{mciteplus.sty has not been loaded}
{This bibstyle requires the use of the mciteplus package.}\fi
\providecommand{\href}[2]{#2}

\newpage
\centerline{\large\bf LHCb collaboration}
\begin{flushleft}
\small
R.~Aaij$^{41}$,
B.~Adeva$^{37}$,
M.~Adinolfi$^{46}$,
A.~Affolder$^{52}$,
Z.~Ajaltouni$^{5}$,
S.~Akar$^{6}$,
J.~Albrecht$^{9}$,
F.~Alessio$^{38}$,
M.~Alexander$^{51}$,
S.~Ali$^{41}$
% ...

{\footnotesize \it
% ...
$ ^{5}$Clermont Universit\'{e}, Universit\'{e} Blaise Pascal, CNRS/IN2P3, LPC, Clermont-Ferrand, France\\
$ ^{6}$CPPM, Aix-Marseille Universit\'{e}, CNRS/IN2P3, Marseille, France\\
% ...
$ ^{9}$Fakult\"{a}t Physik, Technische Universit\"{a}t Dortmund, Dortmund, Germany\\
% ...
$ ^{37}$Universidad de Santiago de Compostela, Santiago de Compostela, Spain\\
$ ^{38}$European Organization for Nuclear Research (CERN), Geneva, Switzerland\\
% ...
$ ^{41}$Nikhef National Institute for Subatomic Physics, Amsterdam, The Netherlands\\
% ...
$ ^{46}$H.H. Wills Physics Laboratory, University of Bristol, Bristol, United Kingdom\\
% ...
$ ^{51}$School of Physics and Astronomy, University of Glasgow, Glasgow, United Kingdom\\
$ ^{52}$Oliver Lodge Laboratory, University of Liverpool, Liverpool, United Kingdom\\
% ...
}
\end{flushleft}


\begin{mcitethebibliography}{10}
\mciteSetBstSublistMode{n}
\mciteSetBstMaxWidthForm{subitem}{\alph{mcitesubitemcount})}
\mciteSetBstSublistLabelBeginEnd{\mcitemaxwidthsubitemform\space}
{\relax}{\relax}

\bibitem{Aubert:2003fg}
BaBar collaboration, B.~Aubert {\em et~al.},
  \ifthenelse{\boolean{articletitles}}{\emph{{Observation of a narrow meson
  decaying to $D_s^+ \pi^0$ at a mass of $2.32 {\,Ge\kern -0.1em V\!/}c^2$}},
  }{}\href{http://dx.doi.org/10.1103/PhysRevLett.90.242001}{Phys.\ Rev.\ Lett.\
   \textbf{90} (2003) 242001}, \href{http://arxiv.org/abs/hep-ex/0304021}{{\tt
  arXiv:hep-ex/0304021}}\relax
\mciteBstWouldAddEndPuncttrue
\mciteSetBstMidEndSepPunct{\mcitedefaultmidpunct}
{\mcitedefaultendpunct}{\mcitedefaultseppunct}\relax
\EndOfBibitem
\bibitem{Besson:2003cp}
CLEO collaboration, D.~Besson {\em et~al.},
  \ifthenelse{\boolean{articletitles}}{\emph{{Observation of a narrow resonance
  of mass $2.46 {\,Ge\kern -0.1em V\!/}c^2$ decaying to $D^{*+}_s \pi^0$ and
  confirmation of the $D^*_{sJ}(2317)$ state}},
  }{}\href{http://dx.doi.org/10.1103/PhysRevD.68.032002}{Phys.\ Rev.\
  \textbf{D68} (2003) 032002}, Erratum
  \href{http://dx.doi.org/10.1103/PhysRevD.75.119908}{ibid.\   \textbf{D75}
  (2007) 119908}, \href{http://arxiv.org/abs/hep-ex/0305100}{{\tt
  arXiv:hep-ex/0305100}}\relax
\mciteBstWouldAddEndPuncttrue
\mciteSetBstMidEndSepPunct{\mcitedefaultmidpunct}
{\mcitedefaultendpunct}{\mcitedefaultseppunct}\relax
\EndOfBibitem
\bibitem{Aubert:2006mh}
BaBar collaboration, B.~Aubert {\em et~al.},
  \ifthenelse{\boolean{articletitles}}{\emph{{Observation of a new $D_s$ meson
  decaying to $DK$ at a mass of $2.86 {\,Ge\kern -0.1em V\!/}c^2$}},
  }{}\href{http://dx.doi.org/10.1103/PhysRevLett.97.222001}{Phys.\ Rev.\ Lett.\
   \textbf{97} (2006) 222001}, \href{http://arxiv.org/abs/hep-ex/0607082}{{\tt
  arXiv:hep-ex/0607082}}\relax
\mciteBstWouldAddEndPuncttrue
\mciteSetBstMidEndSepPunct{\mcitedefaultmidpunct}
{\mcitedefaultendpunct}{\mcitedefaultseppunct}\relax
\EndOfBibitem
\bibitem{Brodzicka:2007aa}
Belle collaboration, J.~Brodzicka {\em et~al.},
  \ifthenelse{\boolean{articletitles}}{\emph{{Observation of a new $D_{sJ}$
  meson in $\Bp \to \Dzb\Dz\Kp$ decays}},
  }{}\href{http://dx.doi.org/10.1103/PhysRevLett.100.092001}{Phys.\ Rev.\
  Lett.\  \textbf{100} (2008) 092001},
  \href{http://arxiv.org/abs/0707.3491}{{\tt arXiv:0707.3491}}\relax
\mciteBstWouldAddEndPuncttrue
\mciteSetBstMidEndSepPunct{\mcitedefaultmidpunct}
{\mcitedefaultendpunct}{\mcitedefaultseppunct}\relax
\EndOfBibitem
\bibitem{Aubert:2009ah}
BaBar collaboration, B.~Aubert {\em et~al.},
  \ifthenelse{\boolean{articletitles}}{\emph{{Study of $D_{sJ}$ decays to
  $D^{*}K$ in inclusive $e^{+}e^{-}$ interactions}},
  }{}\href{http://dx.doi.org/10.1103/PhysRevD.80.092003}{Phys.\ Rev.\
  \textbf{D80} (2009) 092003}, \href{http://arxiv.org/abs/0908.0806}{{\tt
  arXiv:0908.0806}}\relax
\mciteBstWouldAddEndPuncttrue
\mciteSetBstMidEndSepPunct{\mcitedefaultmidpunct}
{\mcitedefaultendpunct}{\mcitedefaultseppunct}\relax
\EndOfBibitem
\bibitem{LHCb-PAPER-2012-016}
LHCb collaboration, R.~Aaij {\em et~al.},
  \ifthenelse{\boolean{articletitles}}{\emph{{Study of $D_{sJ}$ decays to
  $D^+K^0_S$ and $D^0K^+$ final states in $pp$ collisions}},
  }{}\href{http://dx.doi.org/10.1007/JHEP10(2012)151}{JHEP \textbf{10} (2012)
  151}, \href{http://arxiv.org/abs/1207.6016}{{\tt arXiv:1207.6016}}\relax
\mciteBstWouldAddEndPuncttrue
\mciteSetBstMidEndSepPunct{\mcitedefaultmidpunct}
{\mcitedefaultendpunct}{\mcitedefaultseppunct}\relax
\EndOfBibitem
\bibitem{Swanson:2006st}
E.~S. Swanson, \ifthenelse{\boolean{articletitles}}{\emph{{The new heavy
  mesons: a status report}},
  }{}\href{http://dx.doi.org/10.1016/j.physrep.2006.04.003}{Phys.\ Rept.\
  \textbf{429} (2006) 243}, \href{http://arxiv.org/abs/hep-ph/0601110}{{\tt
  arXiv:hep-ph/0601110}}\relax
\mciteBstWouldAddEndPuncttrue
\mciteSetBstMidEndSepPunct{\mcitedefaultmidpunct}
{\mcitedefaultendpunct}{\mcitedefaultseppunct}\relax
\EndOfBibitem
\bibitem{Rosner:2006jz}
J.~L. Rosner, \ifthenelse{\boolean{articletitles}}{\emph{{Hadron spectroscopy:
  theory and experiment}},
  }{}\href{http://dx.doi.org/10.1088/0954-3899/34/7/S07}{J.\ Phys.\
  \textbf{G34} (2007) S127}, \href{http://arxiv.org/abs/hep-ph/0609195}{{\tt
  arXiv:hep-ph/0609195}}\relax
\mciteBstWouldAddEndPuncttrue
\mciteSetBstMidEndSepPunct{\mcitedefaultmidpunct}
{\mcitedefaultendpunct}{\mcitedefaultseppunct}\relax
\EndOfBibitem
\bibitem{Klempt:2007cp}
E.~Klempt and A.~Zaitsev,
  \ifthenelse{\boolean{articletitles}}{\emph{{Glueballs, hybrids, multiquarks:
  experimental facts versus QCD inspired concepts}},
  }{}\href{http://dx.doi.org/10.1016/j.physrep.2007.07.006}{Phys.\ Rept.\
  \textbf{454} (2007) 1}, \href{http://arxiv.org/abs/0708.4016}{{\tt
  arXiv:0708.4016}}\relax
\mciteBstWouldAddEndPuncttrue
\mciteSetBstMidEndSepPunct{\mcitedefaultmidpunct}
{\mcitedefaultendpunct}{\mcitedefaultseppunct}\relax
\EndOfBibitem
\bibitem{Colangelo:2012xi}
P.~Colangelo, F.~De~Fazio, F.~Giannuzzi, and S.~Nicotri,
  \ifthenelse{\boolean{articletitles}}{\emph{{New meson spectroscopy with open
  charm and beauty}},
  }{}\href{http://dx.doi.org/10.1103/PhysRevD.86.054024}{Phys.\ Rev.\
  \textbf{D86} (2012) 054024}, \href{http://arxiv.org/abs/1207.6940}{{\tt
  arXiv:1207.6940}}\relax
\mciteBstWouldAddEndPuncttrue
\mciteSetBstMidEndSepPunct{\mcitedefaultmidpunct}
{\mcitedefaultendpunct}{\mcitedefaultseppunct}\relax
\EndOfBibitem
\bibitem{Wagner:1974gw}
F.~Wagner, M.~Tabak, and D.~M. Chew,
  \ifthenelse{\boolean{articletitles}}{\emph{{An amplitude analysis for the
  reaction $\pi^+ \proton \to \pi^+ \pi^- \pi^0 \Delta^{++}$ at $7 {\,Ge\kern
  -0.1em V}$}}, }{}\href{http://dx.doi.org/10.1016/0370-2693(75)90637-1}{Phys.\
  Lett.\  \textbf{B58} (1975) 201}\relax
\mciteBstWouldAddEndPuncttrue
\mciteSetBstMidEndSepPunct{\mcitedefaultmidpunct}
{\mcitedefaultendpunct}{\mcitedefaultseppunct}\relax
\EndOfBibitem
\bibitem{Aston:1988rf}
D.~Aston {\em et~al.}, \ifthenelse{\boolean{articletitles}}{\emph{{Spin parity
  determination of the $\phi_{J}(1850)$ from $K^- p$ interactions at $11
  {\,Ge\kern -0.1em V\!/}c$}},
  }{}\href{http://dx.doi.org/10.1016/0370-2693(88)90439-X}{Phys.\ Lett.\
  \textbf{B208} (1988) 324}\relax
\mciteBstWouldAddEndPuncttrue
\mciteSetBstMidEndSepPunct{\mcitedefaultmidpunct}
{\mcitedefaultendpunct}{\mcitedefaultseppunct}\relax
\EndOfBibitem
\bibitem{Brandenburg:1975ft}
G.~W. Brandenburg {\em et~al.},
  \ifthenelse{\boolean{articletitles}}{\emph{{Determination of the $K^*(1800)$
  spin parity}},
  }{}\href{http://dx.doi.org/10.1016/0370-2693(76)90711-5}{Phys.\ Lett.\
  \textbf{B60} (1976) 478}\relax
\mciteBstWouldAddEndPuncttrue
\mciteSetBstMidEndSepPunct{\mcitedefaultmidpunct}
{\mcitedefaultendpunct}{\mcitedefaultseppunct}\relax
\EndOfBibitem
\bibitem{Baldi:1976ua}
R.~Baldi {\em et~al.}, \ifthenelse{\boolean{articletitles}}{\emph{{Observation
  of the $K^*(1780)$ in the reaction $K^+ \proton \to K^0_S \pi^+ \proton$ at
  $10 {\,Ge\kern -0.1em V\!/}c$}},
  }{}\href{http://dx.doi.org/10.1016/0370-2693(76)90279-3}{Phys.\ Lett.\
  \textbf{B63} (1976) 344}\relax
\mciteBstWouldAddEndPuncttrue
\mciteSetBstMidEndSepPunct{\mcitedefaultmidpunct}
{\mcitedefaultendpunct}{\mcitedefaultseppunct}\relax
\EndOfBibitem
\bibitem{blatt-weisskopf}
J.~Blatt and V.~E. Weisskopf, {\em Theoretical nuclear physics}, J. Wiley (New
  York), 1952\relax
\mciteBstWouldAddEndPuncttrue
\mciteSetBstMidEndSepPunct{\mcitedefaultmidpunct}
{\mcitedefaultendpunct}{\mcitedefaultseppunct}\relax
\EndOfBibitem
\bibitem{Dalitz:1953cp}
R.~H. Dalitz, \ifthenelse{\boolean{articletitles}}{\emph{{On the analysis of
  tau-meson data and the nature of the tau-meson}},
  }{}\href{http://dx.doi.org/10.1080/14786441008520365}{Phil.\ Mag.\
  \textbf{44} (1953) 1068}\relax
\mciteBstWouldAddEndPuncttrue
\mciteSetBstMidEndSepPunct{\mcitedefaultmidpunct}
{\mcitedefaultendpunct}{\mcitedefaultseppunct}\relax
\EndOfBibitem
\bibitem{LHCb-PAPER-2014-036}
LHCb collaboration, R.~Aaij {\em et~al.},
  \ifthenelse{\boolean{articletitles}}{\emph{{Dalitz plot analysis of
  $B^0_s\to\bar{D}^0K^-\pi^+$ decays}},
  }{}\href{http://dx.doi.org/10.1103/PhysRevD.90.072003}{Phys.\ Rev.\
  \textbf{D90} (2014) 072003}, \href{http://arxiv.org/abs/1407.7712}{{\tt
  arXiv:1407.7712}}\relax
\mciteBstWouldAddEndPuncttrue
\mciteSetBstMidEndSepPunct{\mcitedefaultmidpunct}
{\mcitedefaultendpunct}{\mcitedefaultseppunct}\relax
\EndOfBibitem
\bibitem{Alves:2008zz}
LHCb collaboration, A.~A. Alves~Jr.\ {\em et~al.},
  \ifthenelse{\boolean{articletitles}}{\emph{{The \lhcb detector at the LHC}},
  }{}\href{http://dx.doi.org/10.1088/1748-0221/3/08/S08005}{JINST \textbf{3}
  (2008) S08005}\relax
\mciteBstWouldAddEndPuncttrue
\mciteSetBstMidEndSepPunct{\mcitedefaultmidpunct}
{\mcitedefaultendpunct}{\mcitedefaultseppunct}\relax
\EndOfBibitem
\bibitem{LHCb-DP-2012-004}
R.~Aaij {\em et~al.}, \ifthenelse{\boolean{articletitles}}{\emph{{The \lhcb
  trigger and its performance in 2011}},
  }{}\href{http://dx.doi.org/10.1088/1748-0221/8/04/P04022}{JINST \textbf{8}
  (2013) P04022}, \href{http://arxiv.org/abs/1211.3055}{{\tt
  arXiv:1211.3055}}\relax
\mciteBstWouldAddEndPuncttrue
\mciteSetBstMidEndSepPunct{\mcitedefaultmidpunct}
{\mcitedefaultendpunct}{\mcitedefaultseppunct}\relax
\EndOfBibitem
\bibitem{LHCb-PAPER-2012-056}
LHCb collaboration, R.~Aaij {\em et~al.},
  \ifthenelse{\boolean{articletitles}}{\emph{{Search for the decay $B^0_s \to
  D^{*\mp}\pi^\pm$}},
  }{}\href{http://dx.doi.org/10.1103/PhysRevD.87.071101}{Phys.\ Rev.\
  \textbf{D87} (2013) 071101(R)}, \href{http://arxiv.org/abs/1302.6446}{{\tt
  arXiv:1302.6446}}\relax
\mciteBstWouldAddEndPuncttrue
\mciteSetBstMidEndSepPunct{\mcitedefaultmidpunct}
{\mcitedefaultendpunct}{\mcitedefaultseppunct}\relax
\EndOfBibitem
\bibitem{LHCb-PAPER-2013-022}
LHCb collaboration, R.~Aaij {\em et~al.},
  \ifthenelse{\boolean{articletitles}}{\emph{{Measurements of the branching
  fractions of the decays $B^0_s \to \bar{D}^0 K^- \pi^+$ and $B^0 \to
  \bar{D}^0 K^+ \pi^-$}},
  }{}\href{http://dx.doi.org/10.1103/PhysRevD.87.112009}{Phys.\ Rev.\
  \textbf{D87} (2013) 112009}, \href{http://arxiv.org/abs/1304.6317}{{\tt
  arXiv:1304.6317}}\relax
\mciteBstWouldAddEndPuncttrue
\mciteSetBstMidEndSepPunct{\mcitedefaultmidpunct}
{\mcitedefaultendpunct}{\mcitedefaultseppunct}\relax
\EndOfBibitem
\bibitem{Feindt:2006pm}
M.~Feindt and U.~Kerzel, \ifthenelse{\boolean{articletitles}}{\emph{{The
  NeuroBayes neural network package}},
  }{}\href{http://dx.doi.org/10.1016/j.nima.2005.11.166}{Nucl.\ Instrum.\
  Meth.\  \textbf{A559} (2006) 190}\relax
\mciteBstWouldAddEndPuncttrue
\mciteSetBstMidEndSepPunct{\mcitedefaultmidpunct}
{\mcitedefaultendpunct}{\mcitedefaultseppunct}\relax
\EndOfBibitem
\bibitem{Pivk:2004ty}
M.~Pivk and F.~R. Le~Diberder,
  \ifthenelse{\boolean{articletitles}}{\emph{{sPlot: a statistical tool to
  unfold data distributions}},
  }{}\href{http://dx.doi.org/10.1016/j.nima.2005.08.106}{Nucl.\ Instrum.\
  Meth.\  \textbf{A555} (2005) 356},
  \href{http://arxiv.org/abs/physics/0402083}{{\tt
  arXiv:physics/0402083}}\relax
\mciteBstWouldAddEndPuncttrue
\mciteSetBstMidEndSepPunct{\mcitedefaultmidpunct}
{\mcitedefaultendpunct}{\mcitedefaultseppunct}\relax
\EndOfBibitem
\bibitem{LHCb-PAPER-2012-025}
LHCb collaboration, R.~Aaij {\em et~al.},
  \ifthenelse{\boolean{articletitles}}{\emph{{First evidence for the
  annihilation decay mode $B^+ \to D_s^+\phi$}},
  }{}\href{http://dx.doi.org/10.1007/JHEP02(2013)043}{JHEP \textbf{02} (2013)
  043}, \href{http://arxiv.org/abs/1210.1089}{{\tt arXiv:1210.1089}}\relax
\mciteBstWouldAddEndPuncttrue
\mciteSetBstMidEndSepPunct{\mcitedefaultmidpunct}
{\mcitedefaultendpunct}{\mcitedefaultseppunct}\relax
\EndOfBibitem
\bibitem{LHCb-PAPER-2012-050}
LHCb collaboration, R.~Aaij {\em et~al.},
  \ifthenelse{\boolean{articletitles}}{\emph{{First observations of
  $\bar{B}^0_s \to D^+D^-$, $D_s^+D^-$ and $D^0\bar{D}^0$ decays}},
  }{}\href{http://dx.doi.org/10.1103/PhysRevD.87.092007}{Phys.\ Rev.\
  \textbf{D87} (2013) 092007}, \href{http://arxiv.org/abs/1302.5854}{{\tt
  arXiv:1302.5854}}\relax
\mciteBstWouldAddEndPuncttrue
\mciteSetBstMidEndSepPunct{\mcitedefaultmidpunct}
{\mcitedefaultendpunct}{\mcitedefaultseppunct}\relax
\EndOfBibitem
\bibitem{LHCb-PAPER-2012-048}
LHCb collaboration, R.~Aaij {\em et~al.},
  \ifthenelse{\boolean{articletitles}}{\emph{{Measurements of the
  $\Lambda_b^0$, $\Xi_b^-$, and $\Omega_b^-$ baryon masses}},
  }{}\href{http://dx.doi.org/10.1103/PhysRevLett.110.182001}{Phys.\ Rev.\
  Lett.\  \textbf{110} (2013) 182001},
  \href{http://arxiv.org/abs/1302.1072}{{\tt arXiv:1302.1072}}\relax
\mciteBstWouldAddEndPuncttrue
\mciteSetBstMidEndSepPunct{\mcitedefaultmidpunct}
{\mcitedefaultendpunct}{\mcitedefaultseppunct}\relax
\EndOfBibitem
\bibitem{LHCB-PAPER-2013-011}
LHCb collaboration, R.~Aaij {\em et~al.},
  \ifthenelse{\boolean{articletitles}}{\emph{{Precision measurement of $D$
  meson mass differences}},
  }{}\href{http://dx.doi.org/10.1007/JHEP06(2013)065}{JHEP \textbf{06} (2013)
  065}, \href{http://arxiv.org/abs/1304.6865}{{\tt arXiv:1304.6865}}\relax
\mciteBstWouldAddEndPuncttrue
\mciteSetBstMidEndSepPunct{\mcitedefaultmidpunct}
{\mcitedefaultendpunct}{\mcitedefaultseppunct}\relax
\EndOfBibitem
\bibitem{PDG2012}
Particle Data Group, J.~Beringer {\em et~al.},
  \ifthenelse{\boolean{articletitles}}{\emph{{\href{http://pdg.lbl.gov/}{Review
  of particle physics}}},
  }{}\href{http://dx.doi.org/10.1103/PhysRevD.86.010001}{Phys.\ Rev.\
  \textbf{D86} (2012) 010001}, {and 2013 partial update for the 2014
  edition}\relax
\mciteBstWouldAddEndPuncttrue
\mciteSetBstMidEndSepPunct{\mcitedefaultmidpunct}
{\mcitedefaultendpunct}{\mcitedefaultseppunct}\relax
\EndOfBibitem
\bibitem{Hulsbergen:2005pu}
W.~D. Hulsbergen, \ifthenelse{\boolean{articletitles}}{\emph{{Decay chain
  fitting with a Kalman filter}},
  }{}\href{http://dx.doi.org/10.1016/j.nima.2005.06.078}{Nucl.\ Instrum.\
  Meth.\  \textbf{A552} (2005) 566},
  \href{http://arxiv.org/abs/physics/0503191}{{\tt
  arXiv:physics/0503191}}\relax
\mciteBstWouldAddEndPuncttrue
\mciteSetBstMidEndSepPunct{\mcitedefaultmidpunct}
{\mcitedefaultendpunct}{\mcitedefaultseppunct}\relax
\EndOfBibitem
\bibitem{LHCb-PAPER-2013-056}
LHCb collaboration, R.~Aaij {\em et~al.},
  \ifthenelse{\boolean{articletitles}}{\emph{{Study of beauty baryon decays to
  $D^0 p h^-$ and $\Lambda_c^+ h^-$ final states}},
  }{}\href{http://dx.doi.org/10.1103/PhysRevD.89.032001}{Phys.\ Rev.\
  \textbf{D89} (2014) 032001}, \href{http://arxiv.org/abs/1311.4823}{{\tt
  arXiv:1311.4823}}\relax
\mciteBstWouldAddEndPuncttrue
\mciteSetBstMidEndSepPunct{\mcitedefaultmidpunct}
{\mcitedefaultendpunct}{\mcitedefaultseppunct}\relax
\EndOfBibitem
\bibitem{Skwarnicki:1986xj}
T.~Skwarnicki, {\em {A study of the radiative cascade transitions between the
  Upsilon-prime and Upsilon resonances}}, PhD thesis, Institute of Nuclear
  Physics, Krakow, 1986,
  {\href{http://inspirehep.net/record/230779/files/230779.pdf}{DESY-F31-86-02}}\relax
\mciteBstWouldAddEndPuncttrue
\mciteSetBstMidEndSepPunct{\mcitedefaultmidpunct}
{\mcitedefaultendpunct}{\mcitedefaultseppunct}\relax
\EndOfBibitem
\bibitem{Fleming:1964zz}
G.~N. Fleming, \ifthenelse{\boolean{articletitles}}{\emph{{Recoupling effects
  in the isobar model. 1. General formalism for three-pion scattering}},
  }{}\href{http://dx.doi.org/10.1103/PhysRev.135.B551}{Phys.\ Rev.\
  \textbf{135} (1964) B551}\relax
\mciteBstWouldAddEndPuncttrue
\mciteSetBstMidEndSepPunct{\mcitedefaultmidpunct}
{\mcitedefaultendpunct}{\mcitedefaultseppunct}\relax
\EndOfBibitem
\bibitem{Morgan:1968zza}
D.~Morgan, \ifthenelse{\boolean{articletitles}}{\emph{{Phenomenological
  analysis of $I=\frac{1}{2}$ single-pion production processes in the energy
  range 500 to 700 MeV}},
  }{}\href{http://dx.doi.org/10.1103/PhysRev.166.1731}{Phys.\ Rev.\
  \textbf{166} (1968) 1731}\relax
\mciteBstWouldAddEndPuncttrue
\mciteSetBstMidEndSepPunct{\mcitedefaultmidpunct}
{\mcitedefaultendpunct}{\mcitedefaultseppunct}\relax
\EndOfBibitem
\bibitem{Herndon:1973yn}
D.~Herndon, P.~Soding, and R.~J. Cashmore,
  \ifthenelse{\boolean{articletitles}}{\emph{{Generalised isobar model
  formalism}}, }{}\href{http://dx.doi.org/10.1103/PhysRevD.11.3165}{Phys.\
  Rev.\  \textbf{D11} (1975) 3165}\relax
\mciteBstWouldAddEndPuncttrue
\mciteSetBstMidEndSepPunct{\mcitedefaultmidpunct}
{\mcitedefaultendpunct}{\mcitedefaultseppunct}\relax
\EndOfBibitem
\bibitem{lass}
LASS collaboration, D.~Aston {\em et~al.},
  \ifthenelse{\boolean{articletitles}}{\emph{{A study of $K^- \pi^+$ scattering
  in the reaction $K^- p \to K^- \pi^+ n$ at $11 {\,Ge\kern -0.1em V\!/}c$}},
  }{}\href{http://dx.doi.org/10.1016/0550-3213(88)90028-4}{Nucl.\ Phys.\
  \textbf{B296} (1988) 493}\relax
\mciteBstWouldAddEndPuncttrue
\mciteSetBstMidEndSepPunct{\mcitedefaultmidpunct}
{\mcitedefaultendpunct}{\mcitedefaultseppunct}\relax
\EndOfBibitem
\bibitem{Zemach:1963bc}
C.~Zemach, \ifthenelse{\boolean{articletitles}}{\emph{{Three-pion decays of
  unstable particles}},
  }{}\href{http://dx.doi.org/10.1103/PhysRev.133.B1201}{Phys.\ Rev.\
  \textbf{133} (1964) B1201}\relax
\mciteBstWouldAddEndPuncttrue
\mciteSetBstMidEndSepPunct{\mcitedefaultmidpunct}
{\mcitedefaultendpunct}{\mcitedefaultseppunct}\relax
\EndOfBibitem
\bibitem{Zemach:1968zz}
C.~Zemach, \ifthenelse{\boolean{articletitles}}{\emph{{Use of angular-momentum
  tensors}}, }{}\href{http://dx.doi.org/10.1103/PhysRev.140.B97}{Phys.\ Rev.\
  \textbf{140} (1965) B97}\relax
\mciteBstWouldAddEndPuncttrue
\mciteSetBstMidEndSepPunct{\mcitedefaultmidpunct}
{\mcitedefaultendpunct}{\mcitedefaultseppunct}\relax
\EndOfBibitem
\end{mcitethebibliography}
\end{document}